\documentclass{ws-jaic}
\usepackage{ws-jaic-thm}        % comment this line when `amsthm / theorem / ntheorem` package is used
\usepackage[square]{natbib}

\begin{document}

\markboth{Yoshiyuki Ohmura, Yasuo Kuniyoshi}{Questions for Theories of Consciousness}

\catchline{0}{0}{0000}{}{}

\title{Minimal Set of Questions for Theories of Consciousness: Toward a Unified Explanatory Framework}

\author{Yoshiyuki Ohmura\footnote{corresponding author}, Yasuo Kuniyoshi}

\address{Department of mechano-informatics, The University of Tokyo, 7-3-1, Hongo, Bunkyo-ku, Tokyo, 113-8656, Japan \\
*ohmura@isi.imi.i.u-tokyo.ac.jp}

\maketitle

\pub{Received Day Month Year}{Revised Day Month Year}

\begin{abstract}
A central challenge in consciousness research is the lack of agreement on what a theory of consciousness should explain, which makes it difficult to compare existing theories. We propose a framework for organizing explanatory targets of theories based on a minimal set of seven questions designed to be theoretically neutral, causally and functionally relevant, and applicable across different systems. We focus particularly on the role of causation based on the argument that causal relations cannot be fully specified within standard physical descriptions alone. Introducing an asymmetric causal structure allows internal mechanisms to be represented explicitly and helps distinguish between variable- and structure-level causation. As an example, we apply the proposed framework to analyzing the Dual-Laws Model. The aim of the framework is not to propose a definitive theory but to provide a common basis for analyzing and developing theories of consciousness.

\end{abstract}

\section{Introduction}	
Although many theories of consciousness have been proposed (\cite{seth, kuhn}), they are difficult to compare because they often focus on different aspects of consciousness (\cite{doerig, sattin, signorelli}). Recent efforts to compare theories directly such as through adversarial collaborations (\cite{consortium}) have faced difficulties partly because such theories were designed to explain distinct explanatory targets. One proposed approach to addressing this difficulty has been to evaluate theories of consciousness in terms of shared requirements or criteria (\cite{doerig, delpin2021, percy2025}). For example, \cite{doerig} proposed a set of criteria aimed at testing empirical theories of consciousness. \cite{percy2025}  made a broader attempt to survey and categorize such requirements. While such lists provide useful insights into observed features of consciousness, they are not always formulated as explanatory targets for theories. Thus, there is still no clear consensus on what a theory of consciousness is expected to explain nor on how such explanatory targets should be organized. A further complication arises from the central role often assigned to the ``hard problem'' of consciousness: how subjective experience arises from physical systems (\cite{chalmers}). While this problem is undoubtedly important, focusing on it in isolation presents substantial difficulties for scientific investigation because of its subjective nature and limited observability. Moreover, an exclusive focus on the hard problem risks narrowing the scope of inquiry and overlooking other aspects of consciousness that are more directly related to observable phenomena, functional roles, or causal structures.

Here, we take a different approach. Rather than focusing on a single defining problem, we ask what a theory of consciousness should explain to provide a comprehensive and scientifically tractable account. We propose a minimal set of questions as a working framework for organizing the explanatory targets of theories of consciousness. These questions are not intended to be exhaustive but are instead designed to capture aspects of consciousness that 
\begin{itemize}
\item can be related to experimentally accessible phenomena or theoretically well-defined structures,
\item are relevant to the functional and causal roles attributed to consciousness, and
\item are applicable to both biological and artificial systems.
\end{itemize}
Importantly, not all explanatory targets correspond to intrinsic properties of consciousness. For example, a theory must clarify the relation between the experiencing subject and the initiator of action (self) as well as the causal mechanisms through which conscious processes influence system behavior. The inclusion of these elements reflects a broader view that goes beyond the characterization of conscious contents alone and highlights a key difference between listing properties of consciousness and specifying explanatory targets for theories. While lists of properties (e.g., integration, exclusivity, or seriality) capture important features of conscious contents, they do not by themselves exhaust what a theory of consciousness is expected to explain. Based on these considerations, our set of questions aims to cover key dimensions along which theories of consciousness differ, which include phenomenal aspects, functional roles, causal relations, structural organization, and applicability across systems. By addressing these questions jointly, we aim to provide a common framework that facilitates comparison, integration, and development of theories of consciousness. As a concrete example, we briefly present an application of the framework to a representative model. However, the primary contribution is not to propose a new theory of consciousness but to clarify the range of explanatory targets that any such theory should address.

\section{Proposed framework}
To select the questions that form our framework, we adopted the following set of guiding criteria:

\begin{enumerate}
\item Where possible, the questions should be connected to empirically accessible phenomena or theoretically well-defined structures. This should ensure that the framework remains compatible with scientific investigation and should allow for experimental validation. However, we acknowledge that not all aspects of consciousness are equally amenable to direct empirical treatment. In particular, the ``hard problem'' of subjective experience plays a central role in philosophical and scientific discussions but remains difficult to test directly. Rather than exclude this aspect, we treated it as a constraint: any viable theory should address how subjective experience is accounted for even if it cannot be fully captured by current empirical methods. Thus, empirical accessibility is treated as a guiding principle rather than an absolute requirement.

\item The questions should be theoretically neutral. They should not presuppose a particular theoretical framework such as functionalism, specific neuroscientific models, or representational accounts, which is essential for enabling meaningful comparison across diverse approaches.

\item The questions should be causally or functionally relevant. That is, they should address how consciousness relates to the behavior, organization, or dynamics of a system beyond mere correlations.

\item The questions should be generalizable across different types of systems. In other words, they should be applicable not only to biological systems but also to artificial or computational systems to support a broader understanding of consciousness.

\item The set of questions should be as small as possible while still capturing the core explanatory demands placed on theories of consciousness to reflect the goal of identifying a minimal working framework rather than an exhaustive taxonomy.

\end{enumerate}

In addition, we adopted a conservative approach regarding distinctions that are not yet theoretically established. For example, although the distinction between conscious and unconscious processing has been widely discussed (\cite{baars, cleeremans2014, dehaene2006, percy2026}),  it is often defined in theory-dependent terms. We therefore did not treat it as a primitive assumption. 
Similarly, we did not include challenges that arise from specific theoretical commitments (\cite{doerig, doerig2, emilsson2022})  as general requirements.
Pathological or atypical cases, in which this correspondence breaks down, may provide important insights but are not treated as primary targets in the present framework.

Based on these considerations, the following set of questions is proposed to reflect key dimensions along which theories of consciousness differ.

\begin{enumerate}
\item \textbf{Phenomenal aspect}:
How can subjective experience be accounted for in physical or computational systems?
\item \textbf{Self}: Why does the experiencing subject coincide with the initiator of action?
\item \textbf{Causation}: Does consciousness have causal efficacy within a system beyond mere correlations or predictability?
\item \textbf{State}: How can differences in levels or states of consciousness be explained?
\item \textbf{Function}: What functional or cognitive roles are associated with consciousness?
\item \textbf{Contents}: How can the diversity, structure, and organization of conscious contents be explained?
\item \textbf{Universality}: Can the theory be applied across different types of systems including artificial systems?
\end{enumerate}

Many properties discussed in previous work can be understood as instances or subcases of these categories. For example, features such as conscious processing can be related to the question of function while properties such as binding, integration, valence, and seriality can be understood as aspects of the question of contents. The proposed questions aim to capture the structure of explanatory targets rather than to enumerate all phenomena individually.

\paragraph{Phenomenal aspect}
This question corresponds to the hard problem of consciousness (\cite{chalmers}). 
It is generally difficult to verify whether a system is experiencing anything through external observation, so a theory that focuses solely on this question is unlikely to be scientifically testable (\cite{kleiner}). However, this does not imply that the hard problem can be ignored. Rather, it should be treated as a constraint on theories of consciousness: any viable theory should account for how subjective experience arises even if direct empirical validation remains limited. The generative mechanism of consciousness requires an account of causal structure rather than mere correlations  (\cite{white}). Thus, a theory must clarify the causal basis of consciousness without relying on assumptions such as panpsychism (\cite{lamme2, doerig}).

\paragraph{Self}
This question concerns the relationship between the experiencing subject and initiator of action. A theory must explain why the experiencing subject and initiator are one and the same. Typically, when the experiencing subject and initiator do not coincide (e.g., ``I'' vs ``my body''), neurological disorders are suspected (\cite{ouwersloot, borrelli}), which suggests that coinciding is a crucial property of consciousness. However, this property has not been emphasized in many conventional theories. The coincidence between the experiencing subject and initiator can also be viewed as an epistemological problem: why the experiencing subject recognizes itself as the initiator of action. Nevertheless, theories positing identity between the experiencing subject and initiator are simpler and more intuitive than assuming such metacognition. Some scientific theories assume a homunculus (\cite{baddeley, haggard}),  but this does not resolve the issue. A theory needs to explains the identity of the experiencing subject and initiator of action without resorting to infinite regress.

\paragraph{Causation}
The problem of mental causation is central to the mind–body problem. Because mental processes supervene on physical systems, a key question is whether such supervenient entities can exert a causal influence over physical processes. Although Kim's  argument against supervenient causation and interlevel causation has long dominated the philosophy of mind (\cite{kim}), it assumes a single supervenient entity and cannot be generalized to deny more general interlevel causation.
Here, it is important to distinguish between predictability and causality (\cite{sanchez}). In nonlinear dynamical systems, coarse-grained states often exhibit behaviors that appear to be autonomous or as if they exhibit interlevel causation
(\cite{seth2, rosas}). However, this type of apparent interlevel causation measured by predictability cannot be considered true causality, which can be defined as a cause that does not influence an effect without a causal transmission mechanism (\cite{salmon, pearl2009}).
Explaining the causality of consciousness is crucial for understanding its functional significance.

\paragraph{State}
Consciousness has different levels or states that lack conscious awareness or experience such as non-rapid eye movement sleep and anesthesia. The brain can autonomously alter states of consciousness (\cite{scammell, eban}), which strongly suggests that there is a mechanism for controlling these states. Theories must account for differences in states of consciousness and the internal mechanisms that control them. This has motivated the view that consciousness depends primarily on internal system dynamics although this remains a subject of debate (\cite{morch}).
Integrated information theory (\cite{tononi}) posits that these states are measurable. However, it is still possible that states of consciousness may change in a binary fashion but appear pseudo-continuous owing to the frequency of these transitions. Therefore, we do not consider measurability (\cite{seth, mcfaddent2023}) 
to be a theory-neutral assumption. \cite{seth} proposed the term ``global states'' because they cannot assume perfect order along a single axis, but this is controversial. At the very least, theories of consciousness must account for differences in states of consciousness and its controllability.

\paragraph{Function}
Many studies have investigated the functional roles associated with conscious awareness often by comparing performances with and without awareness. Their results have suggested that consciousness is associated with processes such as semantic integration (\cite{mudrik2, moors}), association (\cite{skora}), and cognitive control (\cite{ansorge, pedale}). 
However, the interpretation of these findings remains debated. For example, it is unclear why certain types of processing appear to require conscious awareness in humans while similar processes can be implemented in artificial systems without it
(\cite{searle, bishop, worden}). 
Therefore, theories of consciousness must explain not only which functions are associated with consciousness but also why such functions depend on consciousness in biological systems.

\paragraph{Contents}
Conscious contents are diverse, structured, and often semantically integrated. For example, in binocular rivalry (\cite{blake}), the experienced percept alternates despite constant sensory input, which indicates a dynamic organization of conscious content. In addition to diversity, theories must account for how conscious contents are integrated into coherent representations (\cite{emilsson2023}) and how they are selectively segmented. Attention plays a key role in this process by influencing which contents become dominant or accessible (\cite{noudoost}). Theories of consciousness must therefore explain not only the variety of contents but also their organization, integration, selection, and seriality (\cite{tononi, mcfaddent2023}). 
In artificial systems, this requires clarifying the relationship between the mechanisms generating physical correlates of consciousness and the resulting structure of contents.

\paragraph{Universality}
Universality is often considered a theoretical virtue (\cite{keas})
and has been proposed as an important requirement for theories of consciousness
(\cite{kanai}). 
Here, universality is treated as a requirement for the explanation rather than as a commitment to substrate independence. Specifically, a theory of consciousness should be applicable across different types of systems provided that the relevant causal and functional structures are present. This does not presuppose that all systems can support consciousness, nor does it assume that implementation details are irrelevant. Rather, it requires that the theory be formulated in a way that allows systematic comparison across different kinds of systems.

\paragraph{Summary}
Existing theories of consciousness often focus on specific aspects of the phenomenon and are not intended to provide a comprehensive account (\cite{sattin, seth, storm, mudrik}). As a result, they are difficult to compare and integrate, which reflects the broader lack of consensus on what a theory of consciousness should explain. The proposed set of questions are not intended to be exhaustive but rather to establish a minimal framework for capturing core aspects of consciousness that any viable theory should address. In particular, the framework emphasizes empirical relevance, theoretical neutrality, causal and functional significance, and applicability across different systems. Importantly, these questions are not intended to be addressed independently in a purely additive manner. A theory of consciousness is expected to provide a coherent and internally consistent explanatory structure that can answer multiple questions through shared mechanisms or principles (\cite{keas}). 
This type of interconnected explanatory structure is characteristic of scientific theories and is an important criterion for evaluating their coherence and predictive power.

\section{An illustrative example}
As an example, we apply the proposed framework to analyzing the Dual-Laws Model (DLM)
(\cite{ohmura}). 
We focus on the question of causation, which is a critical aspect for theories of consciousness as it directly concerns whether consciousness can have a genuine influence on system behavior rather than being treated as a mere epiphenomenon. The DLM aims to explain consciousness from the perspective of causal structures and mechanisms rather than neural correlates or phenomenological characteristics, which makes it a suitable example for illustrating how the question of causation can be formulated and addressed.

\subsection{Causation in theories of consciousness}
The role of causation in theories of consciousness remains conceptually unclear. While many theories appeal to causal relations between mental and physical processes, the status of such causation is often left unspecified. Standard physical theories describe systems in terms of dynamical laws governing the evolution of states over time. However, these laws do not themselves define asymmetry between a cause and effect (\cite{pearl2009}) and instead provide symmetric relations between states without specifying which variables are causes and which are effects. This distinction has important implications for debates on mental causation.
Kim's philosophical arguments against mental causation (\cite{kim}) have often been interpreted as suggesting that causal relations at higher levels cannot be meaningfully introduced within scientific theories. However, such conclusions rely on relatively strong assumptions, which include the identification of physical determinism with causal closure and the treatment of supervenience in terms of a single higher-level entity. As a result, they do not necessarily generalize to models that involve multiple supervenient entities organized in hierarchical structures. The limitations of these arguments suggest that the question of causation remains open to further exploration in theories of consciousness rather than being excluded in principle (\cite{ohmura_cs}). It remains an open question whether causal efficacy can be directly assessed within purely physical descriptions or whether an additional conceptual structure is required. 

\subsection{Distinguishing causal relations from physical descriptions}
Rather than attempt to derive causal relations directly from physical laws alone, an alternative approach is to distinguish between different levels or types of descriptions (\cite{ohmura_cs}). 
This type of approach is consistent with contemporary theories of causation such as Pearl’s structural causal models (\cite{pearl2009}), where causal relations are explicitly defined through asymmetric assignment operators rather than being derived from underlying physical laws. Such an approach allows for the introduction of a causal structure without requiring it to be explicitly encoded in the underlying physical dynamics. In addition, it opens the possibility that causal relations may arise at levels that are not captured by relations between states in physical models. In particular, it is useful for separating physical descriptions based on dynamical laws from causal relations that introduce asymmetric relations between variables. 

Physical descriptions characterize systems according to state-transition laws that relate variables across time. These relations are typically symmetric in that the same equations can often be used to infer past and future states, so they do not by themselves determine a direction of causation. In contrast, causal relations introduce an explicit asymmetry between cause and effect (\cite{pearl2009, woodward}) that is typically grounded in the notion of intervention, where manipulating one variable leads to changes in another while the reverse manipulation does not necessarily hold. Once these roles are separated, apparent conflicts between physical descriptions and causal relations can be reformulated rather than treated as contradictions.

Importantly, the present framework does not commit to a particular ontological stance regarding the distinction between mental and physical processes. Instead, this distinction is understood as arising within a given descriptive framework.
In causal descriptions, higher-level structures can be identified with mental aspects and lower-level dynamics with physical aspects, allowing the formulation of mental causation as a relation between levels. However, the same system can also be described purely in terms of physical dynamics by omitting causal asymmetry, resulting in a unified dynamical model.
From this perspective, whether a system admits a ``mental–physical'' distinction depends on the descriptive framework being adopted, rather than on an underlying ontological division. This makes it possible to analyze mental causation without introducing independent  ontological domains, and to relate such descriptions to physical models in a consistent manner.

A further step is to consider how causal relations themselves should be defined. In many standard formulations, causation is defined at the variable level as in expressions with the form $b := F(a)$, where changes in the variable $a$ are treated as the cause of changes in $b$ while the function $F$ is treated as fixed. 
However, this formulation assumes that the mapping between variables remains unchanged. We must consider the possibility that changes in the mapping itself can also play a causal role if they introduce asymmetric effects that cannot be fully captured by variable-level causal relations alone. From this perspective, it is useful to distinguish between variable-level causes corresponding to changes in input variables and structure-level causes corresponding to changes in the mapping that governs the relationship between variables. This distinction allows causation to be defined not only in terms of changes in system states but also in terms of changes in the structures that organize those states. Such a formulation is particularly relevant for models that involve hierarchical organization, where higher-level dynamics can modify the structure of lower-level processes. In these cases, treating structural changes as causes provides a way to characterize interlevel causation that cannot be captured by variable-level causal relations alone.

\subsection{The Dual-Laws Model}
The DLM introduces a hierarchical system with two levels. The lower level describes state-transition dynamics, and the higher level governs structural modifications of these dynamics. Each level is characterized by its own dynamical laws, but causal relations arise through their interaction. Thus, the DLM explicitly distinguishes between dynamical evolution and causal structure. Causal relations are defined in terms of asymmetric dependency under intervention. In particular, the DLM allows for causal influences that originate from changes in higher-level structures rather than only from changes in lower-level state variables and thus reflects the distinction between variable- and structure-level causes.

This structure can be made more explicit by a simple formal representation. In the DLM, the feedback error can be expressed as  $err := e_c(d)$, where $d$ denotes lower -level input variables and  $e_c$ is a function constructed from a family of supervenient functions ${X^i}_{(i \in I)}$. The structure of $e_c$ is determined by an index sequence $c$ at the higher level. $I \subset \mathbb{N}$ represents a set of indices.  In this formulation, changes in $d$ correspond to variable-level causes. In contrast, changes in $c$ modify the function $e_c$ itself by altering the selection and organization of supervenient functions and therefore can be interpreted as structure-level causes. Crucially, these structure-level causes do not directly act on the lower-level states but influence them through a feedback process. 
The error $err$ generated by  $e_c(d)$ drives changes in the lower-level variables via a feedback control mechanism. The feedback error acts on subvenient entities such as neurons and synapses that constitute the supervenient functions, which modify their states to reduce the error. Because the relationship between the supervenient functions and subvenient entities is constitutive, this mechanism can be interpreted as a form of whole-to-parts causation  (\cite{ohmura_wp}):  changes at the level of the supervenient function influence the behavior of its physical constituents. Importantly, both levels share the same physical entities, and the causal influence is realized within a single integrated system rather than through interactions between separate subsystems.

This hierarchical organization also highlights a distinctive feature of the DLM. The index sequence $c$, which determines the structure of the causal transmission function $e_c$, is not directly determined by the subvenient states that realize those functions. Conversely, changes in subvenient states do not uniquely determine the index sequence. As a result, the higher-level structure and lower-level dynamics can exhibit partially independent dynamics. This separation is important because it allows for the possibility of independent interventions at different levels. Changes in the index sequence $c$ can be treated as a higher-level intervention leading to systematic changes in the function $e_c$ without being reducible to changes in lower-level state variables. Such independence is a key condition for defining causal relations in terms of intervention.
Because  $e_c$  can be modified independently of the lower-level states, it becomes possible to introduce an asymmetric relation between cause and effect, which is formalized by the assignment operator ``$:=$''.

From this perspective, the introduction of causal asymmetry is not merely a formal choice but a way of making the internal mechanisms of the hierarchical system explicit. By distinguishing between changes in variables and changes in structure, the DLM provides a transparent account of how causal influence is generated and transmitted rather than treating these mechanisms as a black box. In contrast, if the same system is described purely as a dynamical model without distinguishing causal asymmetry, the structure of hierarchical causation becomes difficult to identify. The resulting description collapses into a single-level dynamical system in which differences between levels are absorbed into state variables and evolution over time and obscures the organization of causal relations across levels.

\subsection{Addressing the questions of the proposed framework}
We now briefly examine how the DLM answers the set of questions to highlight how the proposed framework can serve as a common structure for analyzing different theories of consciousness.

\paragraph {Phenomenal aspect: }
The DLM associates the generation of consciousness with the causal influence between supervenient functions and lower-level dynamics. This suggests a structural condition under which a system may support consciousness, but the nature of subjective experience itself is not directly explained.

\paragraph{Self:}
The higher-level dynamics that determine the index sequence $c$ can be interpreted as specifying the structure that organizes both action and internal processing. In this sense, the same structural component underlies both the initiation of action and the system’s internal organization, which offers a possible account for the coincidence between the experiencing subject and initiator of action.

\paragraph{Causation:}
The DLM explicitly represents causal efficacy through structure-level causes and their transmission to lower-level states via feedback mechanisms. Causal influence is not reduced to correlations but is formulated in terms of asymmetric relations under intervention.

\paragraph{State:}
The dependence of system behavior on the operation of feedback mechanisms suggests a distinction between different modes of operation, which can be related to differences in states of consciousness.

\paragraph{Function:}  
The interaction between higher-level structures and lower-level dynamics provides a basis for understanding how functional processes depend on system organization. Functions are not defined solely by input–output relations but by internally structured causal mechanisms.

This also suggests that processes accompanied by conscious awareness may differ from similar processes without awareness, not merely in terms of performance or output, but in terms of the presence of internally structured causal mechanisms.

\paragraph{Contents:}
The composition of the function $e_c$ through the selection and organization of supervenient functions provides a way to relate structural organization to the diversity and integration of contents. Different configurations of the index sequence $c$ can give rise to different structured contents within the system. The selection and ordering of supervenient functions may be related to several characteristic features of conscious contents such as their integration into unified representations, their selective activation, and their exclusivity with respect to alternative contents. This suggests a possible link between structural organization and the way conscious contents are organized and experienced.

\paragraph{Universality:}  
Because the DLM is formulated in abstract mathematical terms and does not depend on specific biological details, it can in principle be applied to different types of systems. This illustrates how the requirement of universality can be incorporated at the level of theoretical design.

\paragraph{Summary:}
This example shows how the proposed framework’s emphasis on shared explanatory targets helps clarify the structure of theoretical approaches independently of specific theoretical commitments.

\section{Possible characteristics of theories with explicit causal structure}
The preceding discussion focused on how a theory of consciousness can be structured in terms of explanatory questions and how causal structure can be explicitly incorporated into a model. We now consider what types of system-level characteristics may arise when such a causal framework is adopted. The aim is not to derive necessary conditions specific to the DLM but to identify features that may generally emerge in systems that possess explicit internal causal organization, especially when causal relations extend across hierarchical levels.

When causal relations are explicitly represented within a system across multiple levels, it becomes possible for higher-level structures to influence lower-level dynamics in a systematic manner. This introduces a form of autonomy at the level of system organization. Such autonomy does not depend on a specific modeling choice such as whether the system is described using multiple dynamical levels or as a single dynamical system with emergent properties. Rather, it arises from the presence of structured causal relations that cannot be reduced to purely local or input-driven interactions. The introduction of a causal structure shifts the perspective from systems that passively respond to inputs to systems whose behavior is partly determined by their internal organization.

A possible consequence of such autonomy is partial independence from immediate external inputs. When the system behavior is shaped by internal causal structure, it need not be fully determined by sensory inputs or environmental conditions. This can be interpreted as a form of cognitive decoupling in which internal processes are not strictly tied to ongoing inputs. Although this concept has been discussed in the context of dual-process theories (\cite{stanovic2012,evans2013}),  it can also be understood more generally as resulting from internal causal organization. Cognitive decoupling does not require a specific cognitive architecture but follows from the presence of higher-level structures that can influence system behavior independently of moment-to-moment inputs.

A related implication concerns the possibility of self-determined behavior. If higher-level causal structures influence lower-level processes, they may also play a role in determining how goals or action patterns are generated within the system. Goals need not be fully specified by external inputs or predefined rules but can arise from internal causal dynamics. This introduces a distinction between systems that execute externally defined instructions and systems that can organize their own behavior through internally structured processes. Such a distinction may be particularly relevant in the study of artificial systems, where the ability to generate internally driven behavior is often considered a marker of more advanced or autonomous systems.

These considerations suggest that theories of consciousness that incorporate explicit causal structure may be applicable to the design and interpretation of artificial systems. Systems that exhibit internal causal organization may differ fundamentally from systems characterized solely by input–output mappings. While these features are not sufficient to establish the presence of consciousness, they illustrate how adopting a causal framework can lead to qualitatively different classes of system behavior. In turn, this highlights the importance of clarifying the role of causation in theories of consciousness especially in contexts where artificial systems are involved.

\section{Conclusion}
A fundamental difficulty in consciousness research is the lack of a shared understanding of what a theory of consciousness is expected to explain. Existing theories often focus on specific aspects of consciousness, which makes systematic comparison and integration challenging. To address this issue, we propose a set of questions as a minimal working framework for organizing the explanatory targets of theories of consciousness. These questions are designed to be empirically relevant where possible, theoretically neutral, causally and functionally meaningful, and applicable across different types of systems including artificial ones. We do not claim that addressing all seven questions constitutes a complete theory of consciousness but rather view these questions as minimal requirements for developing a scientifically meaningful and coherent theory. By shifting the focus from individual proposed theories to shared explanatory targets, the proposed framework provides a common basis for comparing, evaluating, and extending theories of consciousness. Future work should explore how different theories address these questions in practice as well as how causal frameworks can be further developed and evaluated, particularly in the context of artificial systems. Clarifying these aspects will be essential for establishing more unified and comprehensive theories of consciousness.

\section*{Acknowledgements}
This research was supported by the JSPS KAKENHI (25H00448), Japan. The funding sources had no role in the decision to publish or prepare the manuscript. 

\section*{Author Contributions}
YO: conceptualization, draft writing, and revision; YK: supervision, funding.

\section*{AI use statement}
AI-assisted tools were used to improve the clarity and readability of the manuscript. However, all ideas, theoretical formulations, and conclusions presented in this work were developed by the authors, who take full responsibility for the content.

\bibliography{sample}
\bibliographystyle{ws-jaic}

\end{document}